\documentclass[aps,prl,twocolumn]{revtex4}
\usepackage{epsf}
\begin{document}

\def\Tr{{\rm Tr}}
\def\Det{{\rm Det}}

\title{Resonating-valence-bond structure of Gutzwiller-projected 
superconducting wave functions}

\author{D.~A.~Ivanov}
\affiliation{Insitute of Theoretical Physics, 
\'Ecole Polytechnique F\'ed\'erale de Lausanne (EPFL), 
1015 Lausanne, Switzerland}

\begin{abstract}
Gutzwiller-projected (GP) wave functions have been widely used for describing
spin-liquid physics in frustrated magnets and in high-temperature
superconductors. Such wave functions are known to represent states of the
resonating-valence-bond (RVB) type. In the present work I discuss
the RVB structure of a GP singlet superconducting
state with nodes in the spectrum. The resulting state for the
undoped spin system may be described in terms of the ``path integral''
over loop coverings of the lattice, thus extending the known 
construction for RVB states. The problem of topological order
in GP states may be reformulated in terms of the statistical behavior
of loops. The simple example of the projected $d$-wave state on 
the square lattice demonstrates that the statistical behavior of 
loops is renormalized in a nontrivial manner by the projection.
\end{abstract}

\maketitle

Gutzwiller-projected (GP) wave functions have been often applied
to describe the unconventional properties of high-temperature
superconductors and the spin-liquid phase of frustrated 
spin-$1/2$ systems \cite{PWA,Gros,Capriotti,vanilla}. 
Since the early days of their use for strongly
correlated systems, GP wave functions have been known not only to
provide a good variational ansatz, but also to represent the
resonating-valence-bond (RVB) physics of the ground state. Because of
their RVB structure, the GP wave functions merit deep investigation as
independent objects, irrespective of the underlying physical Hamiltonian
\cite{toporder1,toporder2}.

It was shown
in the original anaysis of GP superconducting wave functions
\cite{PWA} that such wave functions correspond
to the RVB states with singlet amplitudes $a_{ij}$ given by the
Fourier transform of
\begin{equation}
a(k)=\frac{u_k}{v_k} \, ,
\label{a-uv}
\end{equation}
where $u_k$ and $v_k$ are the coherence factors of the BCS wave function
before projection. For a fully gapped superconductor, this procedure produces
amplitudes $a_{ij}$ decaying exponentially with the site separation $|i-j|$,
and the interpretation as a RVB state is straightforward. However, in the
case of a BCS superconductor with nodes, including the commonly used 
$d_{x^2-y^2}$ state, this simple derivation produces values of $a(k)$ 
singular at 
the nodal points, and the resulting singlet amplitudes $a_{ij}$ are 
long-ranged. Therefore, in the case of a superconductor with nodes, the
conventional derivation cannot interpret the GP wave function as a RVB state.

In this note I resolve the problem of interpreting the physical content
of the GP wave function by using an alternative representation
in terms of the ``loop-soup'' path integral.
In this construction, well-known for RVB states, the correlation
functions are expressed as sums over all close-packed coverings of the
lattice with closed loops \cite{Liang,Sutherland,Kohmoto}. 
In the conventional RVB loop construction,
the statistical weights are determined by the products of singlet
amplitudes along the loops. In my generalized derivation, the product
of singlet amplitudes is replaced by the trace of the product of the BCS
Green's functions along the loop. Thus the role of the RVB singlet amplitudes
is played by the BCS Green's functions (2$\times$2 matrices). In the case
of a superconductor with nodes, the BCS Green's functions are only
marginally local: they decay algebraically with distance. 
It is therefore clear that caution is required when applying the spin-liquid 
RVB scenario to high-temperature superconductors \cite{vison}.

The paper is continued with a brief discussion of the relation of the
loop-soup construction to the topological order predicted for RVB
states \cite{RK,RC,Bonesteel,toporder1,toporder2}. 
The presence of topological order corresponds to disordered
short-ranged loops.

I then employ the example of the undoped GP $d$-wave state on the
square lattice to demonstrate the possibility of strong renormalization
of the loop statistics by projection. The loop-length distribution may be
characterized by spin-spin correlations which decay algebraically with
distance. The power of the algebraic decay is non-universal and depends on
the variational parameter of the wave function.

The derivation begins by defining the GP wave function. I consider 
a system consisting of spins 1/2 occupying a finite lattice
with an even number of sites $N$. The GP wave function is constructed
by using an auxiliary Hamiltonian of the BCS form,
\begin{eqnarray}
H&=&\sum_{\{ij\},\alpha} \left[ t_{ij} \psi^\dagger_{i\alpha} \psi_{j\alpha}
+ {\rm h.c.} \right] \nonumber\\
& & + \sum_{\{ij\}} \left[
\Delta_{ij} (\psi^\dagger_{i\uparrow} \psi^\dagger_{j\downarrow}-
\psi^\dagger_{i\downarrow} \psi^\dagger_{j\uparrow}) + {\rm h.c.}
\right]
\label{hamiltonian}
\end{eqnarray}
(in this general form the
hopping and pairing amplitudes between any pairs of sites
are arbitrary).
The Hamiltonian is spin-rotation invariant (it involves only singlet
pairing), and its ground state $\Psi_{\rm BCS}$ is a spin singlet.
The GP wave function $\Psi_{\rm GP}$ for the undoped system
is constructed by projecting onto
states with exactly one fermion per lattice site
(Gutzwiller projection). The resulting wave function may be written
as a wave function of spin variables $\Psi_{\rm GP}(\{\sigma_i\})$, where
the spins $\sigma_i$ must contain $N/2$ up and $N/2$ down spins.

The Hamiltonian (\ref{hamiltonian}) contains more information than
required for the construction of the wave fucntion. First, the Gutzwiller
projection selects states invariant with respect to SU(2) gauge
rotations in the particle-hole space \cite{Affleck,Zhang}. 
Different
Hamiltonians related by such SU(2) gauge symmetries 
therefore produce
the same projected wave function $\Psi_{\rm GP}$.
Second, the Hamiltonian (\ref{hamiltonian})
contains information not only about its eigenfunctions, but
also about the spectrum, which is not used in the
construction of the wavefunction. To eliminate the latter redundancy,
we shall use the equal-time Green's function (the ``projector operator'')
instead of the Hamiltonian.

To construct the equal-time Green's functions, we first define the
$2N$-dimensional vector space of fermionic operators
\begin{equation}
\gamma=\sum_i \left[
u(i)\psi^\dagger_{i\uparrow}+v(i)\psi_{i\downarrow} \right]
\label{gamma-definition}
\end{equation}
(the sum is taken over the $N$ lattice sites). 
This space has the Hermitian form defined by the anti-commutator
$\{\gamma^\dagger_1,\gamma_2\}$, and the Hamiltonian (\ref{hamiltonian})
acts in this space as a Hermitian operator. The eigenvalue equations
are the Bogoliubov--de Gennes equations
\begin{equation}
[H,\gamma]=E\gamma.
\label{bdg}
\end{equation}
The spectrum of the eigenvalues consists of pairs of opposite
energies $\pm E$: this is the consequence of the spin-rotational
invariance of the Hamiltonian (\ref{hamiltonian}). For each
solution $(u,v)$ with energy $E$, the pair $(v^*,-u^*)$ gives
a solution with energy $-E$. Thus there are $N$ positive-energy
solutions and $N$ negative-energy solutions to Eq.~(\ref{bdg}).

Next we define the projector onto the negative-energy states,
\begin{equation}
G=\sum_{E_k<0} \left| \gamma_k \right\rangle \left\langle \gamma_k \right|,
\end{equation}
where $\left| \gamma_k \right\rangle$ are normalized eigenstates
[solutions of (\ref{bdg})]. The operator $G$ constructed in this
way is a $2N\times 2N$
Hermitian matrix which may be considered as a single-particle
projector onto the negative-energy states or, equivalently,
as the matrix of equal-time Green's functions. It contains
no information about the eigenvalues of (\ref{bdg}), but only
about the eigenfunctions. Thus the projector $G$ 
contains less information than the
original Hamiltonian, but remains sufficient for
constructing the multi-particle ground-state wave function.

In real space, the operator $G$ may be represented by the set of 
$2\times 2$ matrices $G_{ij}$ for each pair of lattice
sites $(i,j)$. The hermiticity implies
\begin{equation}
G_{ij}^\dagger = G_{ji}.
\end{equation}

The SU(2) gauge transformation acts on $G$ by conjugation
\begin{equation}
G_{ij} \mapsto W_i^\dagger G_{ij} W_j,
\end{equation}
where $W_i$ are the SU(2) matrices of the gauge transformation.

From the completeness of the basis of all states $\gamma_k$ and from
the $E\to -E$ symmetry (i.e. from the spin-rotational invariance), 
we can derive the following identity for $G_{ij}$:
\begin{equation}
G_{ij}+J^\dagger G^*_{ij} J = \delta_{ij} {\bf 1},
\qquad J=\pmatrix{0 & 1 \cr -1 & 0}.
\label{P-symmetry}
\end{equation}
Mathematically, this identity expresses the fact that $G-\frac{1}{2}$
belongs to the Lie algebra sp($N$) (see, for example, 
Ref.~\onlinecite{Zirnbauer}).
One may verify that Eq.~(\ref{P-symmetry}) is invariant with respect to
SU(2) gauge rotations.

From the SU(2) invariance of Gutzwiller projection, it follows that
the wave function depends only on the SU(2)-gauge-invariant properties 
of the Green's functions $G_{ij}$. For future application, we define the cyclic
trace of the Green's functions
\begin{equation}
T_{i_1 \dots i_l} = \frac{1}{2}\Tr (G_{i_1 i_2} G_{i_2 i_3} \dots G_{i_l i_1})\, ,
\label{c-trace}
\end{equation}
which depends on the oriented closed loop $i_1 i_2 \dots i_l$. 
This type of cyclic trace appears below in the loop path integral.

To proceed with the loop construction, we first express the
wave function in terms of the negative-energy states $\gamma_k$ and
further develop the loop path integral for ground-state expectation
values.

The BCS ground state of the Hamiltonian (\ref{hamiltonian}) may be
expressed as
\begin{equation}
\Psi_{\rm BCS}
=\prod_{E_k<0} \gamma_k \left| \downarrow...\downarrow \right\rangle,
\end{equation}
where $\left| \downarrow...\downarrow\right\rangle$ is the state with all 
lattice sites occupied by down spins, 
$\left| \downarrow...\downarrow\right\rangle=\prod_i \psi^\dagger_{i\downarrow}
\left| 0 \right\rangle$, and the operators $\gamma_k$ are the negative-energy
eigenvectors of Eq.~(\ref{bdg}).

After projecting onto singly-occupied states, this wave function may be
shown to yield for the GP wave function
\begin{equation}
\Psi_{\rm GP}(\{\sigma_i\})=\Det_{k,i} \left[ u_k(i) | v_k(i) \right],
\label{wave-function-determinant}
\end{equation}
where $[ u_k(i) | v_k(i) ]$ is the $N\times N$ matrix 
whose first $N/2$ columns are composed of the coefficients $u_k(i)$ 
while the last $N/2$ columns involve those of $v_k(i)$. 
The index $k$ referes to all
negative-energy states and the index $i$ to all spin-up sites
in the given spin configuration $\{\sigma_i\}$.

Alternatively, the same wave function may be written in the same form
(\ref{wave-function-determinant}), but with the index $i$ labeling
spin-down sites. It follows from the rotation invariance that these
two determinants must give the same wave function, to within an overall
phase factor.

The experssion for the ``partition function''
$\langle\Psi|\Psi\rangle$ is obtained by taking
for the bra-vector the determinant
(\ref{wave-function-determinant}) over the spin-up sites and for the
ket-vector the same determinant over the spin-down sites. 
The product of the two determinants is then rewritten
as the determinant of
the product of the two matrices (taking the sum over the index $k$)
to yield
\begin{equation}
|\Psi_{\rm GP} (\{\sigma_i\})|^2 = {\rm const}\, \Det_{ij} [G_{ij}],
\label{partition-function}
\end{equation}
where $i$ labels spin-up sites and $j$ labels spin-down sites.
The matrix $[G_{ij}]$ is the $N\times N$ matrix composed of  
$2\times 2$ blocks $G_{ij}$.
The constant prefactor does not depend on the spin configuration
$\{\sigma_i\}$ and will be omitted in further calculations.
The expression (\ref{partition-function}) is explicitly SU(2)
gauge-invariant.

\begin{figure}
\centerline{
\epsfxsize=0.8\hsize
\epsfbox{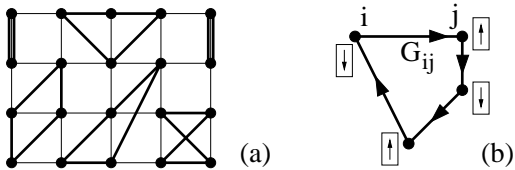}
}
\smallskip
\caption{
{\bf (a)}
 An example of covering the lattice by loops in the ``loop-soup''
construction.
{\bf (b)} To each link of the loop, there corresponds a $2\times 2$ 
matrix Green's function $G_{ij}$.
Each (oriented) loop is assigned the amplitude
$-2T_{i_1...i_l}$. The origin of the negative sign in the loop 
amplitude lies in the
fermionic statistics in the GP construction.
The vertices are labeled with up or down spins which alternate along the loop. 
The factor of two in the loop amplitude is due to the two possible 
spin labelings of the loop.
}
\label{fig1}
\end{figure}

The determinant (\ref{partition-function}) may be further expanded 
as a product of
non-intersecting loops and summed over all possible spin configurations
$\{\sigma_i\}$ to produce the partition function $\langle\Psi|\Psi\rangle$.
Brief algebraic manipulations using the
symmetry (\ref{P-symmetry}) give the loop path integral
\begin{equation}
\langle\Psi|\Psi\rangle=
\sum_{\{C_n\}} 
\prod_n (-2T_{i_1...i_{l_n}})\, ,
\label{partition-function-soup}
\end{equation}
where the sum is taken over all coverings by non-intersecting loops
of even length, and $T_{i_1...i_{l_n}}$ are the loop traces
defined in Eq.~(\ref{c-trace}), see Fig.~\ref{fig1}. The sum
(\ref{partition-function-soup}) [and all loop sums below] is to be
understood as that over {\it oriented} loops. From the symmetry
(\ref{P-symmetry}) it follows that the traces $T_{i_1...i_{l_n}}$
for even-length loops are purely real and do not depend on the 
orientation of the loop. However, it is important to preserve
loop orientations in the definition of the sum (\ref{partition-function-soup}),
because they affect the multiplicities of length-two loops: every
length-two loop has only one orientation and appears in the
sum (\ref{partition-function-soup}) once; all longer loops admit
two orientations and appear twice.

This result has the form of a path integral in which
different correlation functions may be computed. For example, the
spin-spin correlation function $\langle S_z(i) S_z(j) \rangle$ may
be written as
\begin{equation}
\langle S_z(i) S_z(j) \rangle = \frac{1}{\langle\Psi|\Psi\rangle}\,
\sum_{\{C_n\}}^{i\leftrightarrow j} (-1)^{P_{ij}}
\prod_n (-2T_{i_1...i_{l_n}})\, ,
\label{correlation-soup}
\end{equation}
where the sum is now taken only over loop coverings with the sites
$i$ and $j$ belonging to the same loop, and $(-1)^{P_{ij}}$ takes 
values $\pm 1$
depending on whether the number of loop links between $i$ and $j$ is even
or odd.

The path integral (\ref{partition-function-soup}) has the same form as
the loop construction for RVB states. Most generally, the loop
path integral may be written as
\begin{equation}
\langle\Psi|\Psi\rangle=
\sum_{\{C_n\}} 
\prod_n A(C_n)\, ,
\label{general-soup}
\end{equation}
where $A(C_n)$ is an amplitude depending on the geometry
of the loop $C_n$. 

This general formulation includes many different RVB-type 
wave functions in a variety of systems. The GP wave functions
in Anderson's derivation \cite{PWA} correspond to the loop
amplitude
\begin{equation}
A(C_n)= -2 a_{i_1 i_2} a_{i_2 i_3} \dots a_{i_l i_1}\, ,
\label{amplitude-RVB}
\end{equation}
where $a_{ij}$ are the singlet amplitudes defined from (\ref{a-uv}).
Note the importance of the negative sign in the loop amplitudes
(\ref{partition-function-soup}) and (\ref{amplitude-RVB}). This
sign arises from the fermionic statistics involved in the GP
construction. Alternatively, for a spin system it is possible to
construct a ``bosonic'' product of singlets which leads to a
loop amplitude similar to Eq.~(\ref{amplitude-RVB}), but {\it without}
the negative sign \cite{Liang,Sutherland,Kohmoto}.

Remarkably, the Rokhsar--Kivelson ground state of dimer models \cite{RK}
may also be described with the same formalism (\ref{general-soup}):
it is sufficient to set $A(C_n)$ equal to 1 for length-two loops on allowed
dimer positions and to 0 otherwise.

\begin{figure}
\centerline{\epsfxsize=0.8\hsize \epsfbox{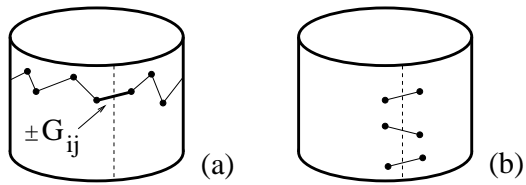}}
\smallskip
\caption{
Defining topological sectors on a cylinder. The ``reference line''
(dashed) connects the two edges of the cylinder. {\bf (a)} The ``plus-minus''
sectors in the loop path integral are generated by changing the sign
of Green's functions $G_{ij}$ intersecting the reference line.
{\bf (b)} The ``even-odd'' sectors in the RVB states are selected as
those with an even (odd, respectively) number of singlets intersecting the
reference line.}
\label{fig2}
\end{figure}

The concept of the ``loop soup'' aids in 
visualizing the conditions required for the topological
order proposed for RVB states \cite{RK,RC,Bonesteel}. 
As explained in Refs.\ \onlinecite{toporder1,toporder2}, the
different topological sectors on a multiply connected domain (e.g. on a 
cylinder or a torus) may be accessed by imposing periodic or antiperiodic
boundary conditions on the fermions in the Hamiltonian (\ref{hamiltonian})
along a topologically nontrivial contour. As a result, the 
periodic/antiperiodic boundary conditions produce two different GP wave
functions $\Psi_+$ and $\Psi_-$. The topological order implies two
dual conditions
\begin{equation}
\langle \Psi_+ | X | \Psi_+ \rangle -
\langle \Psi_- | X | \Psi_- \rangle  \to  0\, ,
\label{toporder-1}
\end{equation}
\begin{equation}
\langle \Psi_+ | X | \Psi_- \rangle \to  0\, ,
\label{toporder-2}
\end{equation}
where the correlation functions must tend to zero sufficiently
rapidly (e.g., exponentially) with increasing system size
for any local operator $X$.

It will become clear below that the first of these 
conditions (\ref{toporder-1})
may be viewed as an effective absence of infinite loops in the path
integral
(\ref{general-soup}), and may be related loosely to the short range
of spin correlations. The second condition (\ref{toporder-2})
corresponds to the absence of valence-bond crystallization in the
RVB construction, and, in the general case (\ref{general-soup}), may
in all probability be formulated as the absence of loop crystallization. 
Note, however, that the following discussion of those correspondences
is not fully rigorous and some statements are indicated as conjectures.

\begin{figure}
\centerline{
\epsfxsize=0.95\hsize
\epsfbox{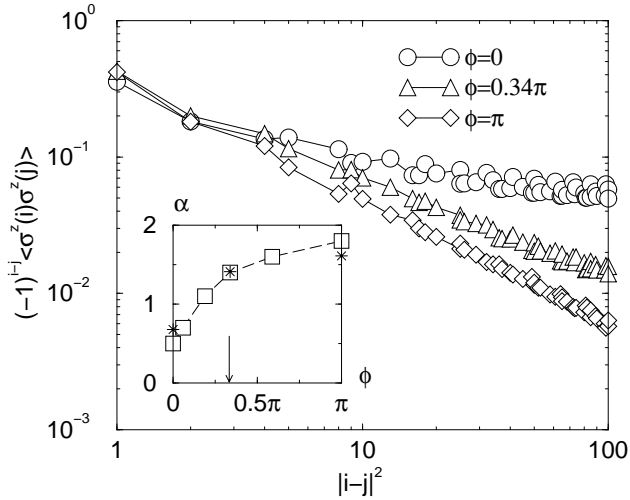}
}
\smallskip
\caption{
{\bf Main panel:} staggered spin correlations in the undoped projected $d$-wave
state for different values of the variational parameter 
$\phi=4\arctan\tilde{\Delta}$. The Monte Carlo simulation was performed on the 
24$\times$24 lattice and involved averaging over $10^4$ samples.
The error bars are smaller than the symbol sizes.
{\bf Inset:} The exponent $\alpha$ as defined in Eq.~(\ref{alpha-def})
as a function of the variational parameter $\phi$. Squares: values of $\alpha$
obtained from the spatial decay of spin correlations in the $24\times 24$ 
system. Stars: values of $\alpha$ obtained from the size-dependence of
the squared integrated staggered magnetization $M(L)$ defined 
in Eq.~(\ref{stag-integral}), see Fig.~\ref{fig4} below.
The error bars
are of the order of the symbol sizes. The arrow marks the
position of the optimal value of the variational parameter
($\phi\approx0.34\pi$) minimizing the variational energy of the
Heisenberg Hamiltonian within the given class of wave functions
\cite{Dmitriev,Yokoyama}.
}
\label{fig3}
\end{figure}

\begin{figure}
\centerline{
\epsfxsize=0.9\hsize
\epsfbox{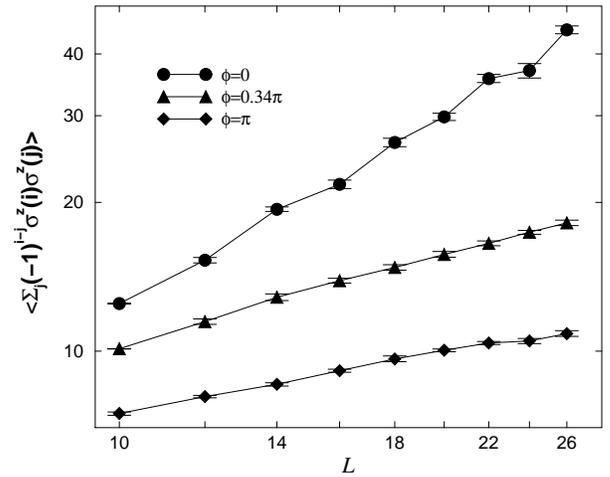}
}
\smallskip
\caption{
The dependence of the fluctuating integrated staggered magnetization $M(L)$
defined in Eq.~(\ref{stag-integral}) on the linear system size $L$
in the log-log scale. The values of $\alpha$ deduced from these
data are plotted by star symbols in the inset of Fig.~\ref{fig3}.
}
\label{fig4}
\end{figure}

I begin by considering the condition (\ref{toporder-1}). 
Under the assumption that that the Green's
functions $G_{ij}$ decay rapidly with distance, loops of
large size may appear only with many links of short length. In this case,
the change of boundary conditions in Eq.~(\ref{partition-function-soup}) may
be represented by changing the sign of the Green functions $G_{ij}$ 
intersecting the reference line (Fig.\ \ref{fig2}). Consequently, the partition
functions in the $\pm$ sectors may be written as
\begin{equation}
\langle\Psi_\pm|\Psi_\pm\rangle=
\sum_{\{C_n\}} (\pm 1)^{W}
\prod_n (-2T_{i_1...i_{l_n}})\, ,
\label{partition-function-topological}
\end{equation}
where $W$ is the total winding number of the loops. This
definition may be extended directly to the general formalism 
(\ref{general-soup}) and to any expectation value of a local observable $X$,
\begin{equation}
\langle\Psi_\pm|X|\Psi_\pm\rangle=
\sum_{\{C_n\}} (\pm 1)^{W} X(\{C_n\})
\prod_n A(C_n)\, .
\label{correlation-topological}
\end{equation}
Therefore the difference between the two topological
sectors (\ref{toporder-1}) contains only configurations with odd winding
numbers. The absence of large loops then serves as a sufficient condition
for the criterion (\ref{toporder-1}) of topological order
[I believe that in most situations this is also a necessary condition]. 
On the other hand, the absence
of large loops also guarantees that spin correlations
(\ref{correlation-soup}) are short-ranged, 
but is not a necessary condition.
Therefore one may expect that in many cases the short range of spin
correlations is related to the condition (\ref{toporder-1}), but
in this contribution I provide no rigorous derivation of such a relation.

We consider next the second condition (\ref{toporder-2}). For the RVB
wave functions represented as linear combinations of products of singlets,
this condition corresponds to the absence of singlet ordering (e.g.
the absence of a valence-bond crystal \cite{Lhuillier}). 
This can be shown by
employing the even-odd basis defined conveniently in the case of RVB or
dimer states on a topologically non-trivial 
domain \cite{Bonesteel,Ioselevich}. Using the
expression (\ref{correlation-topological}) for the definition
of $\Psi_\pm$, the ``plus-minus'' basis is related to the ``even-odd'' 
basis by
\begin{equation}
\Psi_\pm = \Psi_e \pm \Psi_o\, ,
\end{equation}
where $\Psi_o$ and $\Psi_e$ are the states with even and odd
numbers of intersections of the singlets (dimers) with the reference line
(Fig.\ \ref{fig2}). The condition (\ref{toporder-2}) may then be reformulated
as
\begin{equation}
\langle \Psi_e | X | \Psi_e \rangle -
\langle \Psi_o | X | \Psi_o \rangle  \to  0
\end{equation}
which, in turn, can be interpreted as the absence of singlet (dimer) 
crystallization \cite{Ioselevich}. 
One may expect that for the more general loop
construction (\ref{general-soup}) this condition also corresponds to the
absence of loop crystallization. However I do not have a rigorous argument
supporting this conjecture.

As discussed above, the first criterion of the topological order 
(\ref{toporder-1}) involves the question of the effective loop size
in the path-integral ensemble (\ref{general-soup}). This effective
loop size cannot be inferred easily from the singlet amplitudes
$a_{ij}$ (in the case of the RVB construction) or from the
Green's functions $G_{ij}$. Numerical studies of spin correlations
in ``bosonic'' RVB states \cite{Liang} indicate that the effective
loop size may be renormalized nontrivially by the constraint
of fully-packed loops.
Similar renormalizations of loop statistics may also be observed
in the fermionic GP construction. I consider for illustration 
one of the simplest
examples of a GP wave function, the undoped projected $d$-wave state on
the square lattice. This state has been studied previously as an approximate
ground state of the Heisenberg antiferromagnet \cite{Dmitriev,Yokoyama}. 
It does not possess long-ranged antiferromagnetic order, but only a
power-law decay of antiferromagnetic correlations \cite{PhD,toporder2}. 
The only dimensionless
parameter in the wave function is the pairing strength 
$\tilde{\Delta}=\Delta/t$, and the wave function is invariant under
$\tilde{\Delta} \to \tilde{\Delta}^{-1}$ 
(see, for example, Ref. \onlinecite{Yokoyama}). Equivalently,
the same wave function may be defined 
as the projected normal staggered-flux state with
a flux per plaquette of $\phi=4\arctan\tilde{\Delta}$ 
\cite{Affleck,Zhang,Ivanov}.
Numerical Monte Carlo simulations indicate that for all values
of the parameter $\phi$, the antiferromagnetic correlations obey
the power law,
\begin{equation}
(-1)^{i-j} \langle S_z(i) S_z(j)\rangle \propto |i-j|^{-\alpha}\, ,
\label{alpha-def}
\end{equation}
see Fig.~\ref{fig3}. 
One observes that the spin correlations decay
very slowly (with $\alpha<2$), which implies that the effective loop
size in the loop path integral (\ref{general-soup}) is of the
order of the system size. The condition (\ref{toporder-1})
of the topological order is therefore expected to fail. 
Interestingly, in Ref.\  \onlinecite{toporder1} this state was indeed
classified as one without topological order, but on the basis of the
failure of the {\it other} condition (\ref{toporder-2}) [the condition
(\ref{toporder-1}) was not considered for this state].
Note that in the current example (undoped projected $d$-wave state) 
the spin correlations
(\ref{alpha-def}) probe precisely the probability of the two
sites $i$ and $j$ belonging to the same loop: the Green's functions
$G_{ij}$ connect only sites of opposite antiferromagnetic sublattices,
and therefore the term $(-1)^{P_{ij}}$ in 
(\ref{correlation-soup}) is equal to $(-1)^{i-j}$
independently of the loop configuration.

The actual value of the exponent $\alpha$ in the 
power law (\ref{alpha-def}) remains a subject of controversy.
One possible scenario is the universal value of $\alpha$ depending
only on the symmetries of the wave function (for example, the authors of
Ref.\ \onlinecite{Randeria} propose the value $\alpha=3/2$).
However, my numerical results suggest a different possibility:
a parameter-dependent non-universal exponent $\alpha(\phi)$. The
values $\alpha(\phi)$ extracted from the spatial decay of spin correlations
are shown in the inset of Fig.~\ref{fig3}. To verify those results,
I have also computed the square of the
integrated staggered magnetization in the $L\times L$ system,
\begin{equation}
M(L)=\frac{1}{L^2} \langle \left[\sum_i (-1)^i S_z(i)\right]^2 \rangle\, .
\label{stag-integral}
\end{equation}
The dependence of $M(L)$ on the linear system size $L$ is plotted in
Fig.~\ref{fig4} for three values of $\phi$. A simple scaling argument
predicts $M(L)\propto L^{2-\alpha}$. The values of $\alpha(\phi)$ deduced
from $M(L)$ are plotted by star symbols in the inset of Fig.~\ref{fig3}.
While those results are consistent with a $\phi$-dependent value of $\alpha$,
they do not constitute a definite proof: a more detailed analysis of 
finite-size effects or an analytic argument is needed to
settle the issue.

To summarize, the GP superconducting wave functions admit a loop
description generalizing the conventional loop construction for
RVB states. The role of the singlet amplitudes is played by the
equal-time BCS Green's functions. The properties of topological order
and fractionalization are related to the correlations of the loops,
but these cannot be inferred simply from the unprojected Green's 
functions involved. Nevertheless,
the loop formulation may be helpful in visualizing the RVB properties
of the GP wave functions, and possibly for developing appropriate
analytic approximations.

The author thanks G.~Jackeli and B.~Normand 
for helpful comments on the manuscript.

\end{document}